\documentclass[conference]{IEEEtran}

\usepackage{graphicx}
\usepackage{multirow}
\usepackage{array}
\usepackage{subfigure}
\ifCLASSINFOpdf

\else
\usepackage{algorithm}
\usepackage{algorithmic}

\begin{document}
\title{Performance Study of Localization Techniques in Wireless Body Area Sensor Networks}

\author{\IEEEauthorblockN{Obaid ur Rehman, Nadeem Javaid, Ayesha Bibi, $^{\$}$Zahoor Ali Khan\\}

          Department of Electrical Engineering, COMSATS\\ Institute of
        Information Technology, Islamabad, Pakistan \\
        $^{\$}$Faculty of Engineering, Dalhousie University, Halifax, Canada
        }

\maketitle

\begin{abstract}
One of the major issues in Wireless Body Area Sensor Networks (WBASNs) is efficient localization. There are various techniques for indoor and outdoor environments to locate a person. This study evaluating and compares performance of optimization schemes in indoor environments for optimal placement of wireless sensors, where patients can perform their daily activities. In indoor environments, the performance comparison between Distance Vector-Hop algorithm, Ring Overlapping Based on Comparison Received Signal Strength Indicator (ROCRSSI), Particle filtering and Kalman filtering based location tracking techniques, in terms of localization accuracy is estimated. Results show that particle filtering outperforms all. GPS and several techniques based on GSM location tracking schemes are proposed for outdoor environments. Hidden Markov GSM based location tracking scheme efficiently performs among all, in terms of location accuracy and computational overheads.
\end{abstract}
\IEEEpeerreviewmaketitle
\IEEEpeerreviewmaketitle


\section{Introduction}
With development of wireless devices and wireless communication in  medical industry, research on Wireless Body Area Sensor Networks (WBANs) attains significance attention. WBAN consist of large number of sensor nodes with wireless communication interface. Sensors provide a cheaper and effective way to manage and care for patients suffering from illness or in the process of rehabilitation. WBASN is a cheaper technology with the attention of treating a human as patient and providing personal network around human body. These networks consists of low power and noninvasive wireless bio sensors implemented in human body to provide a smart healthcare system. Information from sensors is provided to medical server placed in hospital to treat patients by a concerned person.

The inherent characteristics of these sensor networks make localization an important issue in WBASNs. Localization identifies position of target sensor nodes in a randomly distributed network. To assign measurement for location each node has to determine its own position.

Location tracking is measured through different location schemes. These schemes are classified into range free and range based schemes, as shown in fig.1. Range based schemes receive location information based on Time of Arrival (TOA), Received Signal Strength Indication (RSSI), Time Difference of Arrival (TDOA) and Angle of Arrival (AOA). After determining range information between nodes then estimating their location through these information. Range based schemes earn higher accuracy than range free for location tracking in various environments. However, major drawbacks conclude these information corrupted by noise and fading and requires additional devices for measuring range information. In Range free schemes, unknown nodes use relative connectivity information from anchors for location estimation. These schemes employ range information based on Approximation Point in Triangle (APIT), Centriod and Distance Vector Hop (DV-Hop). Range free schemes not require additional devices for measuring range information, therefore less effected by environmental changes than range based schemes.

In this paper, we analytically survey different localization techniques. These schemes are divided into two categories: indoor environments and outdoor environments. In indoor environments, it is difficult to predict path loss due to multipath and shadowing. Signal effects due to scattering, reflection and diffraction and also effects by changing indoor environments like motion of peoples inside building. In outdoor environments pathloss prediction is easier, because path is mostly line of sight between mobile and source. The implementation and performance methods in indoor and outdoor environments are totally different, as shown in table.1. We discuss: DV-Hop Algorithm, ROCSSI, particle filtering, Kalman filtering location tracking for indoor environments. GSM based Tracking techniques includes Cell ID based, Deterministic fingerprints based, Probabilistic fingerprints based, Hidden Markov Model (HMM) location tracking and GPS based location tracking are proposed for outdoor environments.

In rest of the paper, related work is given in section 2, section 3 discusses indoor location tracking schemes, section 4 describes outdoor location tracking techniques.

\begin{figure*}[ht]
\begin{center}
\centering

\includegraphics [scale=0.6]{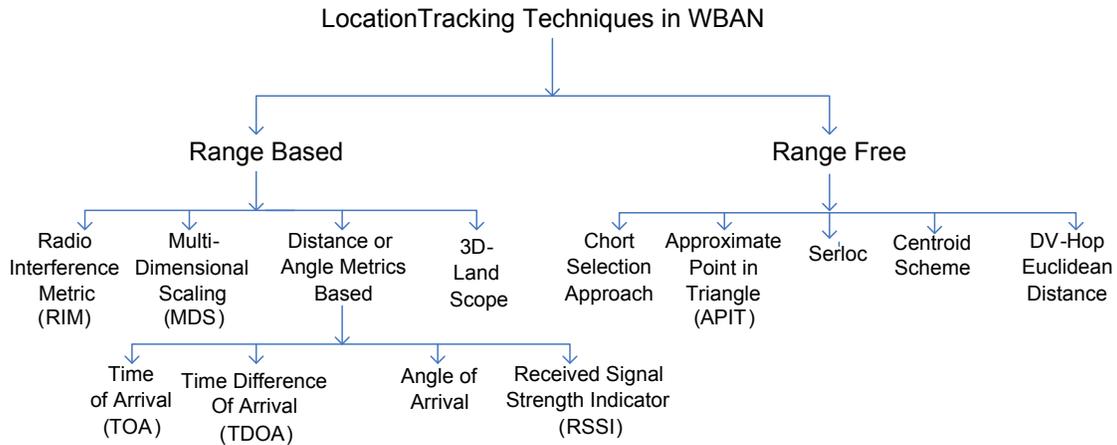}
\caption{Hierarchy of Location Tracking Techniques}
\end{center}
\end{figure*}

\begin{table*}[ht]
\centering
\begin{tabular}{|c |c |c |} 
\multicolumn{3}{c}{Table.1.Comparison of Indoor and Outdoor Location Tracking}\\
\hline
 Features & Outdoor & Indoor \\
\hline                  
  Path loss model & Linear & Affected by Multipath and Shadowing \\
\hline
 Accuracy & Easy to achieve but not necessary & Difficult to achieve but important \\
\hline
 Space& Wide and not limited  & Small and mostly Rectangular \\
\hline
 Deployment & Random and Ad-Hoc & Can be Planned \\
\hline
 Maps & Global & Local \\[1ex]      
\hline 
\end{tabular}
\end{table*}

\section{Related Work}
Need of location tracking in WBASNs is essential for patient moving in indoor and outdoor environments. In recent advancements several localization schemes are adopted, keeping eyes on application requirement and demand to locate a person in body area networks.

GSM based on cellular signals. In GSM, each cell is distinguished by unique cell identifier (ID) and allocated to one or more up/down link frequency pairs. Base Transceiver Station (BTS) acts as an anchor node and mobile node is moving within range of these anchors. Position is determined by taking average of received anchor positions [1].

RSSI based algorithm is used to track location of a person.  Spatial Diversity is used to combat fading effects, a ring is generated when target node is between two anchor nodes (Beacon Nodes). After generating a series of rings, we count number of times area is covered. Gravity is calculated at centre of area to find accurate location [2].

Particle filtering based localization algorithm is used Bayesian Posterior probabilistic distribution method to estimate unknown node location.
Time series location information is expressed by evaluation of particles.
In weighting phase of particle filtering is evaluated by likelihood of the particles. Message overheads are reduced by piggybacking power levels with control messages [3].

In GPS location tracking technique, mobile nodes move in 3D space and periodically broadcast their position information through beacon messages. Static nodes receive beacon messages, when they are in communication range of mobile nodes. The static nodes calculate their position using equation of Sphere [4].

A range free, DV-Hop algorithm scheme is used to track location. Where, anchor nodes broadcast their location information in entire network. An anchor mode receives location information and minimum hop count from other anchors, average hop size can be determine for a single hop. Relative distance is estimated by unknown nodes using hop size and minimum hop count [11].

Several localization tracking methods for GSM location tracking technique are Cell ID based, deterministic fingerprinting based, probabilistic fingerprinting based and Hidden markov model are proposed in [12] [13] [14] .

In indoor localization techniques, we compare results in terms of localization accuracy. For nonlinear systems result shows, particle filtering technique outperforms all. For outdoor environments, we compare results in terms of localization accuracy and computational overheads. We suggests mathematical equations for Deterministic fingerprinting location tracking scheme and HMM location tracking scheme.

\begin{figure*}[!t]
  \centering
\subfigure[Number of Anchor Nodes = 3 , Number of Unknown Nodes = 50]{\includegraphics[height=5 cm,width=9 cm]{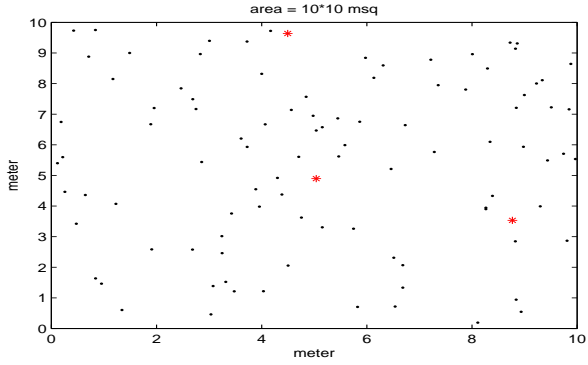}}
\subfigure[Error Estimation in Fig.a]{\includegraphics[height=5 cm,width=9 cm]{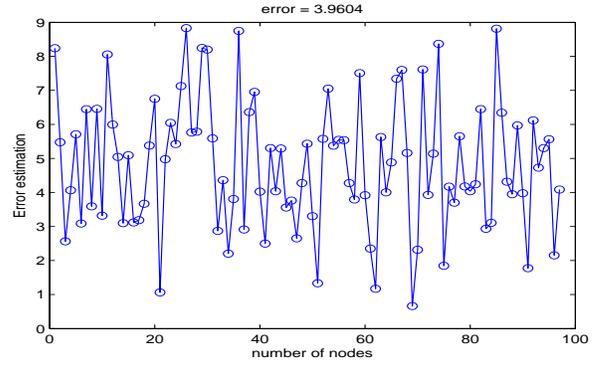}}
\subfigure[Number of Anchor Nodes = 25 , Number of Unknown Nodes = 50]{\includegraphics[height=5 cm,width=9 cm]{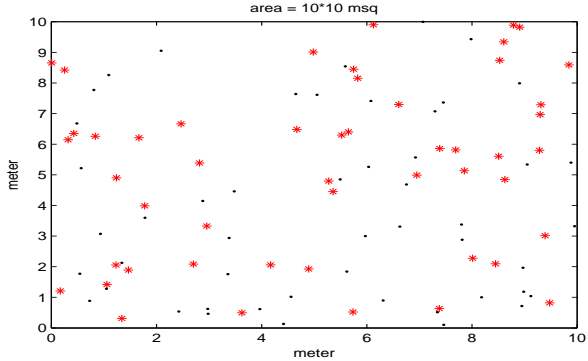}}
\subfigure[Error Estimation in Fig.c]{\includegraphics[height=5 cm,width=9 cm]{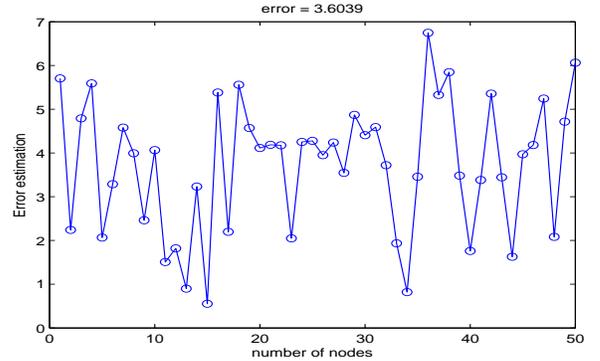}}
\caption{DV-Hop Algorithm for Location Tracking}
\end{figure*}

\section{Indoor Location Tracking Techniques}
In indoor environments, it is difficult to predict path loss. This is because of  multipath and shadowing. Signals are effected due to scattering, reflection, diffraction and  by change in environment, like motion of peoples inside building. In this paper, we discuss four algorithms for indoor location tracking; DV-Hop Algorithm, ROCSSI, particle filtering and Kalman filtering.

\subsection{DV hop Algorithm}
DV-Hop algorithm is based on distance vector. Location of unknown nodes is estimated by calculating distance from anchors, regardless of nodes, which do not have ability to measure the range information. In DV-hop scheme, anchor nodes broadcast their location information in entire network.
Location information and minimum hop count obtained by anchor from other anchors, whereas, average hop size can be determined for a single hop. Relative distance is estimated by an unknown node using hop size and minimum hop count from them.

All simulations are performed in MATLAB. The network area is $10\times10$ square meter, radio nodes having transmission range of 2m. we can roughly estimate the positioning error of the algorithm with the length of line, longer line length represents longer location error. Basic DV-Hop algorithm performs well on the regular uniform topology. Its performance is not efficient in non uniform topology. All simulations results are taken by changing number of anchor nodes, as number anchor nodes increases location error is minimize. However, one drawback increasing the computational overheads by increasing number of anchor nodes.
This scheme not gives accurate results for location tracking.

\begin{algorithm}
\small
\caption{RSSI based location tracking}
\begin{algorithmic}
\STATE $RSSI\; Propagation {}$
\STATE $Number\;of\;samples\leftarrow N $
\FORALL {t=1: T}
\STATE $broadcast\;beacon\;messages {}$
\IF{$t=T $}
\STATE $take\; mean\; of \;sampled \;received \;from\; neighbors$
\STATE $\frac{\sum RSSI}{N}$
\ENDIF
\STATE $Broadcast\;RSSI\;message$
\STATE $RSSI\;message\;=\;Mean \;of\;RSSI\;+\;ID$
\STATE$RSSI\; Estimation \;process$
\STATE $inputs$
\STATE $Assume\; that\; sensor \;node\; ``S``\; Wants\; to \;conclude \;its\; Location$
\STATE $A\;,B\;,C\; denotes \;anchor \;nodes$
\STATE $RSSI_{AB}\;RSSI_{AC}\;RSSI_{AS}\;are \;the \;received\; signal \;strength$
\STATE $Ring\;set\;R\;=\;()$
\IF{$D_{AS}>D_{AB}$and$D_{AS}<D_{AC}$}
\STATE$R_I(A)=D_{AB}$
\STATE$and \;Outer \;Radius\; Will \;equal \;to$
\STATE$R_o(A)=D_{AC}$

\ENDIF
\IF{$RSSI_{AB}>RSSI_{AS}$ and $ RSSI_{AS}>RSSI_{AC}$ $with\;A,\;B,\;S,\;and\;C\;in\;same\;direction\;$}

\STATE$ S\;is\;in\;shaded\;area $
\STATE$ Generate \;a\;ring\;R\;centered \;at\;A\;with\;inner\;radious\;d_1$
\STATE$and \;outer\;radious\;d_2$
\ENDIF
\ENDFOR
\end{algorithmic}
\end{algorithm}

\subsection{ROCSSI Based Location Tracking}
In [2], RSSI based algorithm is used to track the location. RSSI is the function of distance, if the value of RSSI is small, it corresponds to higher estimation error and long distance. To achieve good accuracy due to deficient antenna coverage and multi path interference, spatial diversity is used to combat the fading affects. In RSSI based networks, beacon messages are send periodically. In [7], range free algorithm is used to track relative position to decide the possible target region.

In this process every time a ring is generated when target node is between two anchor nodes (Beacon Nodes). After generating a series of rings, we count the number of times in which area is covered. Every time a ring covers an area, the counter of area is increased by 1. This scheme calculates the gravity center of the area to find the accurate location. In [9] Rings are generated as portray in figure. 3.

\begin{figure}[h]
\tiny
\begin{center}
\includegraphics [scale=0.40]{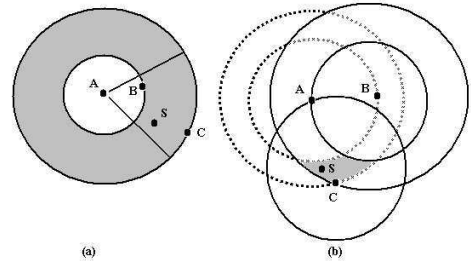}
\caption{ ROCRSSI Based Loaction Tracking }
\end{center}
\end{figure}

\subsection{Location Finding through Particle Filtering}
In [3], particle filtering based localization algorithm is used. Particle filters mostly used in nonlinear systems. This algorithm receives radio signal strength information from beacon messages from its neighbors to infer its position. It uses finite random particles for probability density function sampling. Bayesian Posterior probabilistic distribution method is used to estimate unknown node position. The inference of time series location information is expressed by the evaluation of particles. In the weighting phase of particle filter, we evaluate the likelihood of the particles. The most unlikely particles will be replaced by most likely ones, particles coverage focus a point step by step. we discuss three models for location tracking in particle filtering. These are target model, sensor model and observation model.

\begin{algorithm}
\caption{Target Model}
\small
\begin{algorithmic}
\STATE $Transitional \;Probability \leftarrow states\; p(T)$
\STATE $Probability\; at\; any \;instant \leftarrow states\; p(t)$
\IF{$p_{(t-1)} = 0$ and $p(t) = 1$}
\STATE $P_{(T)}=1$
\STATE $P_{(int)}=1\; means \;target\; is \;present$
\ENDIF
\IF{$P_{(t-1)} = 1$ and $ p_{(t)} = 1$}
\STATE $p_{(int)}=0$ and $p_{(out)} = 1$
\STATE $means\; target \;is \;absent$
\ENDIF
\end{algorithmic}
\end{algorithm}

\subsubsection{Target Model}
The target in the data is modeled as Binary Markov Process. The target presence variable, $P_t$ can take on two values, normally,  $P_t=0$ indicating the absence of target. $P_t=1$, indicates presence of target. At any instant target can present at any point. Disappearance of target means that intensity of target signal strength goes down below Threshold level ($\mu$). We propose transitional probabilities of target initialization and target outage probability $P_{out}$ is modeled as follows:
\begin{equation}
P_{int}=P_r{(P_t=1|P_t-1=0)}
\end{equation}
\begin{equation}
P_{out}=P_r{(P_t=0|P_t-1=1)}
\end{equation}
$P_{int}=1$ will occur when $P_r>\mu$.

\subsubsection{Sensor Model}
In [6], sensor model is describe in detail. System of ``N'' small devices deployed over an area in an attempt to sense a signal transmitted by the target. Placing of anchors can either be regular pattern or deployed in ad-hoc manners. The binary decision is made at each instant ``t'' on the basis of ``M`` samples of received signals. At a particular instant ``t'', each sensor can either be active or inactive. Each active sensor makes a binary decision about whether, target is present or absent.

Energy per sample of a target at  $i_{th}$ sensor in [6] is as:
\begin{equation}
E^2_T(d_i)=E^2_{To}/d_i^2
\end{equation}
where $E^2_{T0}$ is energy per sample of target signal at a distance of 1 unit and $d_i$ denotes distance between target at $i_{th}$ sensor.

Each sensor performs an assumption test between $A_o$ (target absence) and $A_1$ (Target presence) assumption model. $A_o$ indicates that energy received from  target is negligible and target is apart from the sensor. Thus under $A_o$, G is Gaussian Vector, whose elements are independent with variance $\sigma_n^2$. Similarly $A_1$ indicates that energy is received from target is significant and target is closer to the sensor. Under $A_1$, G is Gaussian Vector whose elements are independent Variance $\sigma^2_N+E^2_N(d_i)$  in [6] is as:

\begin{equation}
A_0=\sigma^2_N
\end{equation}

\begin{equation}
A_1=\sigma^2_N+E^2_N(d_i)
\end{equation}

\begin{equation}
P(y:A_1)=\frac{1}{\sqrt{2\pi[\sigma^2_N+E^2_N(d_i)]}}e^{-\frac{y}{2(\sigma^2_N)+E_N^2(d_i)}}
\end{equation}

\begin{equation}
P(y:A_0)=\frac{1}{\sqrt{2\pi\sigma^2_N}}e^{-\frac{y}{2(\sigma^2_N)}}
\end{equation}

We propose conditional probability that event $A_0$ occurs:

\begin{equation}
P(y;A_0|y;A_1)=\frac{P(y;A_1 \times y;A_0)}{P(y:A_1)}
\end{equation}

Applying conditional probability that event $A_1$ occurs:

\begin{equation}
P(y;A_1|y;A_0)=\frac{P(y;A_0 \times y;A_1)}{P(y:A_0)}
\end{equation}

\begin{algorithm}
\small
\caption{Sensor model}
\begin{algorithmic}
\STATE $N$ = number of devices deployed in regular or adhoc manners
\STATE $t \;states \leftarrow time\; instant$
\STATE $d_i states \leftarrow distance \; between \; target \; and \; i_{th} \;sensor$
\STATE $M states \leftarrow Samples \; of \; received \; signals$
\STATE $ Energy\; per \; sample \; of \; transmitted \; signal\leftarrow  E^2_{t_0}$
\STATE Energy received per sample of target at ith sesor is $E^2_N(d_i) = \frac{E^2_{t_0}} {d_i^2} $
\STATE $target \; absence \leftarrow states (A_0)$
\STATE $target\; presence \leftarrow states\; (A_1)$
\IF{$A_1=\sigma^2_N+E^2_N(d_i)$}
\STATE $means \;target \;is \;present$
\ELSE[$A_1=\sigma^2_N$ ]
\STATE $target\; is \;absent$
\ENDIF
\STATE $Applying \;condition\:\emph{} probability$
\STATE $P(y;A_0|y;A_1)=\frac{P(y;A_1 * y;A_0)}{P(y:A_1)}$
\STATE$where$
\STATE$P(y:A_1)=\frac{1}{\sqrt{2\pi[\sigma^2_N+E^2_N(d_i)]}}e^{-\frac{y}{2(\sigma^2_N)+E_N^2(d_i)}}$
\STATE$and$
\STATE$P(y:A_0)=\frac{1}{\sqrt{2\pi\sigma^2_N}}e^{-\frac{y}{2(\sigma^2_N)}}$
\STATE$P(y;A_1|y;A_0)=\frac{P(y;A_0 * y;A_1)}{P(y:A_0)}$
\STATE$P(y;A_1|y;A_0) \leftarrow states \; probability \;of \;A_1\; occurance$
\end{algorithmic}
\end{algorithm}

\subsubsection{Observation Model}
The number of active sensors determines the size of observation vector. Vector $ Z_k $ contains binary observations from each active sensor at a given time $ k $. If target is not detects, corresponding element of $ z_k $ becomes zero, otherwise one.

Probability distribution of single node in [6] is modeled as:

\begin{equation}
p(z_k(i)|x_k) = [P_D(di)]^{z_k(i)}[1- P_D(di)]^{1-z_k(i)}
\end{equation}

The probability distribution of vector $z_k$ in [6] is:

\begin{equation}
 p(z_k(i)|x_k)=\prod_{i=1}^{n}
[PD(di)]^{z_k(i)}[1-PD(di)]^{1-z_k(i)}
\end{equation}
\subsection{Kalman Filtering}

Kalman filters are mostly used in linear systems, however these systems are very few in numbers in world. priori and posterior probability distribution of Kalman filter is Gaussian. Bayesian probability distribution process helps to model the kalman filter. This probability distribution function is discussed below:

\subsubsection{General Bayesian Tracking Model}
Motion of a person modeled using general bayesian tracking model in [11] as follows:

\begin{equation}
x_k = f_k(x_{k-1},u_{k-1},w_k)
\end{equation}

\begin{equation}
z_k=h_k(w_k , u_k , v_k)
\end{equation}

Current location of person modeled by a nonlinear function $ f_k $, which depends on the previous location. $ h_k $ is a nonlinear observation function. Current location of person can be estimated at each step recursively with update and prediction stage.

\subsubsection{Prediction Stage}

In [11], prediction stage is modeled as:

\begin{equation}
p(x_k|z_{1:k-1})=\int p(x_k|x_{k-1})p(x_{k-1}|z_{1:k-1})d_{x_{k-1}}
\end{equation}

\subsubsection{Update Stage}
In [11],m update stage is modeled as:

\begin{equation}
p(x_k|z_{1:k})=\frac{p(z_k|x_k)p(x_k|z_{1:k-1})}{p(z_k|z_{1:k-1})}
\end{equation}

\begin{equation}
p(x_k|z_{1:k-1})=\int p(z_k|x_k)p(x_k|z_{1:k-1})d_{x_k},
\end{equation}

If observation and process noise are assumed to be Gaussian then general filtering reduces to a Kalman filter.
\subsubsection{Kalman filter}

Kalman filters measurement equations in [11] are as follows:

\begin{equation}
x_k=f_kx_{k-1}+w_k , w_k \sim N(0,Q) , X(0) \sim N(X(0),V(0))
\end{equation}

\begin{equation}
Z_K = H_kx_k+v_k ,v_k \sim N(0,R)
\end{equation}

The measurement and process noise are defined covariance matrix $ Q $ and $ R $ and assumed to be independent \\ The prediction and update stage of Kalman filter is given by following equations from [11] is.

\subsubsection{Predict stage for Kalman filter}

\begin{equation}
Z_k^\wedge = F_{x_k-1}^\wedge ,
\end{equation}

\begin{equation}
Z_k^- = F_{p_{k-1}}F^T+Q,
\end{equation}

\subsubsection{Update stage for Kalman filter}
\begin{equation}
k_k = p_k^-H^T(HP_k^-H^T+R)^-1
\end{equation}

\begin{equation}
x_k^\wedge = x_k^-+ K_k(Z_k-H_{xk}^\wedge-)
\end{equation}

\begin{equation}
P_k = (1-k_kH)p_k^-
\end{equation}

Initially current location is predict using previous location. The estimations are updated using weighted observations by the Kalman gain ($k_k$). If the variance is high, process noise variance matrix $ R $ will be large, thus decreases Kalman gain and effects observation. Kalman gain becomes small, if posteriori error variance $p_k$ is low, it  gives more significance to the prediction.

\subsection{Simulation results for particle and kalman filters}
All simulations are performed in the MATLAB. Reason behind selection this tool is necessary matrix operations are implemented to program Particle filter and Kalman Filter.

All simulations are performed for the case, where initial state is known, true state of target is provided to the filter. In our case 50 particles are used and senors are randomly distributed. Figure. 4-a shows probability distribution function at a specific time interval in discrete and continuous manners for particle filter. In figure. 4-b, we estimate RMS error for particle filter and kalman filter. Results shows estimated error for kalman filter and particle filter at specific time step. The estimated error for kalman filter is 0.92679 and estimated error for particle filter is 0.62056. We suggest, in a random and nonlinear systems particle filter is best suited among all of four location tracking techniques discussed for indoor environments. Particle filter is accurate location tracking technique, however implies greater computational overheads is major drawback.

\begin{figure*}[!t]
  \centering
\subfigure[PDF For particle filter at Specific Time]{\includegraphics[height=5 cm,width=9 cm]{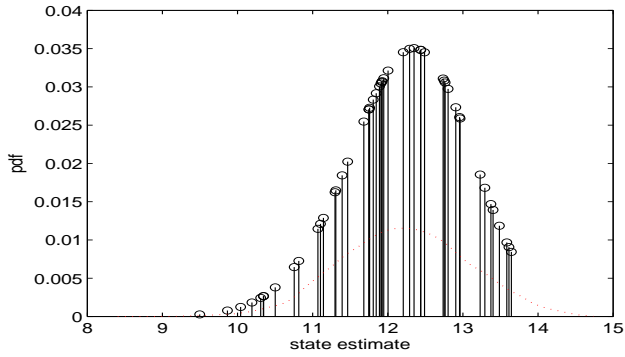}}
\subfigure[Error Estimation Through Particle Filter and Kalman Filter]{\includegraphics[height=5 cm,width=9 cm]{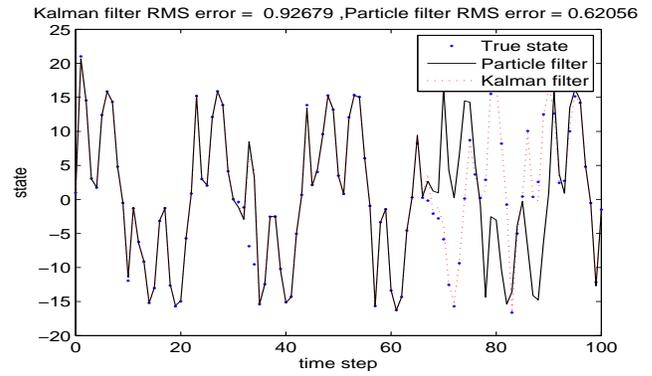}}
\caption{Location Tracking Through Particle and Kalman Filter}
\end{figure*}

\section{Outdoor Location Tracking Techniques}
In outdoor, environments path loss prediction is easier, because path is mostly line of sight between mobile and source. GSM based tracking techniques, includes Cell ID based, Deterministic fingerprints based, Probabilistic fingerprints based, HMM location tracking and GPS based location tracking are proposed for Outdoor environments.

\begin{figure}[ht]
\begin{center}
\includegraphics [scale=0.5]{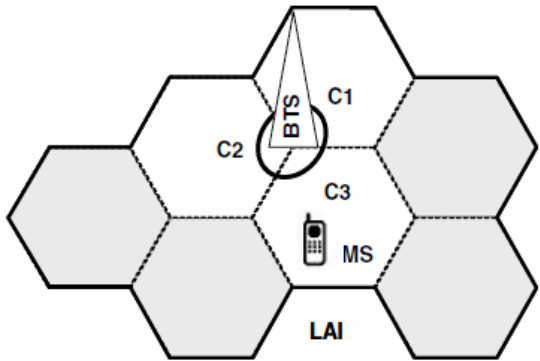}
\caption{ Cell ID based location tracking technique.Mobile station is within cell C3 }
\end{center}
\end{figure}

\subsection{GSM Based Tracking}
In [1], GSM technique is used for location in WBASNs. GSM network is divided into numbers of cells and each cell contains Base Transceiver Station. Numbers of BTS are controlled by Base Station Controller (BSC) and also many BSC's are controlled by many MSC. GSM has licensed band of frequencies and do not suffer from interference. In this scheme all the BTS act as an anchor node and mobile node is moving within the range of these anchors. As soon as a mobile node receives announcements from these anchors, it is able to determine that it is in intersection of the circles centered at these anchors. Position is determined by taking average of received anchor positions.\\GSM based localization is available on cell phones, which presents ``80`` to ``90`` percent. GSM systems are mainly RSSI based and it is a function of distance. GSM localization techniques are further divided in Cell ID based, Deterministic fingerprinting techniques, Probabilistic fingerprinting technique and HMM location tracking techniques.

\subsubsection{Cell ID based Tracking}
Cell ID based positioning does not require any upgarde of network equipments and is simple and economic. Location accuracy depends on cell size. Accuracy of this localization scheme ranges from some meters to kilometers. In cell ID based localization technique BTS covers a number of cells, each cell identified by unique cell ID $ (C_1 , C_2 , C_3)$ in fig. 10. A mobile station (MS) continuously select a cell, exchanges data and signaling traffic with corresponding BTS. Cells are formed in shape of cluster. MS always knows ID of current cluster, each cell in cluster is identified by Local Area Identifier (LAI),  such techniques require database from cell towers.

\subsubsection{Deterministic Fingerprinting  Based Tracking}
In [13], Deterministic Fingerprinting Based Location tracking techniques, this technique stores information about the received RSSI from different based stations at different locations. This process is usually constructed in offline phase. During Location tracking phase, the K-Nearest Neighbors (KNN) algorithm is used, where received RSSI at an unknown position is compared to RSSI signatures in the fingerprint. Closest location in fingerprint returned as estimated location in terms of Euclidian distance in RSSI spaced.\\ We proposed mathematical equation for nearest neighbor method simply calculates the Euclidean distance between current RSSI reading in tracking phase and each reference point fingerprint in offline phase.

\begin{equation}
E_D =\sqrt{ \sum_{i=1}^N(RSSI_T-RSSI_{FP})^2}
\end{equation}

Where $ E_D $ = Euclidean distance between current RSSI reading in tracking phase and reference point fingerprint in offline phase.\\
$RSSI_T$ = RSSI receive in tracking phase.\\
$RSSI_{FP}$ = RSSI received in off line phase.

Deterministic fingerprinting provide higher accuracy, however require searching a large database than cell ID Based technique and constructing a fingerprint also time consuming process.
\subsubsection{Probabilistic Fingerprinting  Based Tracking}
In [13], probabilistic location tracking techniques provide more accurate location. However constructing a probabilistic fingerprint is more complex and challenging. To construct a signal strength histogram for certain amount of time, we need to stand at each fingerprint. This process add significant overhead for fingerprinting construction process. Cell sense addresses this challenge by using gridding. In this approach area of interest divided into number of grids, for each grid a histogram is constructed. This technique removes extra overheads and also helps to reduce the fingerprint size by increasing the grid cell size. This approach works in two phases, offphase and online phase.

Offline phase construct signal strength histogram for RSSI received at each location of fingerprint for each cell. To avoid fingerprint construction overheads, area of interest is divided into cells. The histogram is constructed for each cell tower by using fingerprints locations inside the cell, rather than each fingerprint point in figure . 6.

\begin{figure}[ht]
\begin{center}
\includegraphics [scale=0.6]{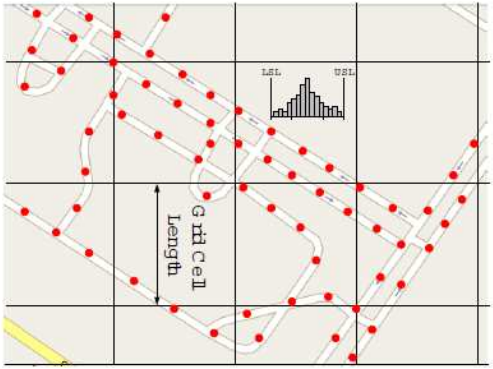}
\caption{ Cellsence approach for figerprint construction }
\end{center}
\end{figure}
\emph{}

During Online phase user is standing at an unknown postion $ ``l'' $ receiving signal strength vector $ s = (s_1, ......,s_q) $, maximum probability to received signal strength $ ``s'' $ from cell tower $ ``i''$ at location $ ``l'' $ is in [13] is modeled as:

\begin{equation}
p(s|l) = \prod_{i=1}^q\prod_{j=1}^Np(s_{i,j}|l)
\end{equation}
where $ s_{i,j} $ represents the $ j^{th} $sample from the $ i^{ith} $ stream.
\subsubsection{Hidden Markov  Model based Tracking}

 HMM location tracking technique proposed in [14]. This technique is based on GSM localization using only RSSI information from associated cell tower. The area of interest is divided into grids, as shown in fig.11. HMM technique include two phases: offline phase and online phase.

Offline phase is used to construct HMM and estimates its parametes. Each state represents a location in the discrete physical space and observation from a state represents the RSSI readings from adjacent cell towers. The parameters of this phase are $ (S,V,A,B,\pi) $. These parameters are estimated in [14] as:

$S$: Each state in model represents a physical grid, $``N''$ number of grid cells and number of states.\\
$V$: At every state, the set of observations related to several pairs that received inside cell.\\
$A$: Estimate the transistion state matrix.\\
$B$: Estimate the observation probability at each state.\\
$\pi$: If initial state is known, it can be used as it is, if any information is not available then steady state probability distribution $ \pi_{ss}$ can be used to estimate the initial state distribution. This distribution estimated by transition probability matrix, $A$ as $ \pi_{ss}A = \pi_{ss}$.

During online phase, user is moving in the area of interest RSSI information is received from the adjacent cell only. History of RSSI values from adjacent cell tower is used to estimate user location. Given the sequence of observations of length $T$, $ O = (O_1,......0_T)$. To find user location at the end of sequence, probable sequence of sates $ Q = (q_1,......,q_t) $ gives sequence of observations. $q_t$ is returned as estimated user location.

\begin{figure*}[ht]
\begin{center}
\centering
\includegraphics [scale=0.55]{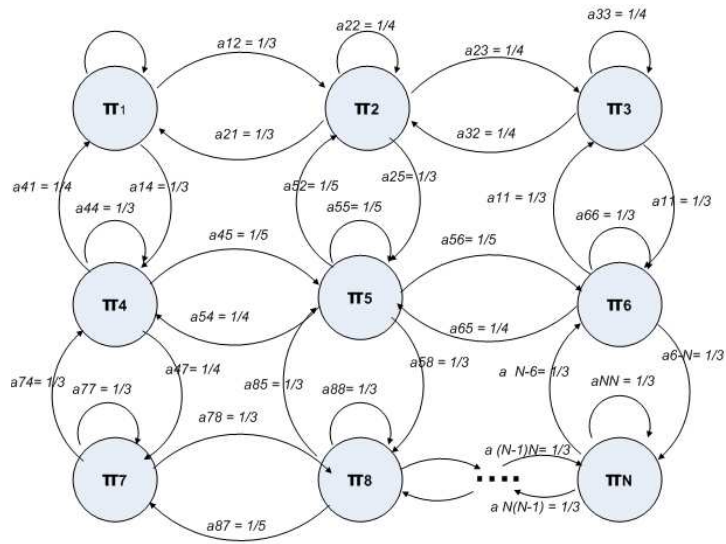}
\caption{ Equivalent HMM for figure.6. Transition Probability For Each State to Its Three Neighboring States Only}
\end{center}
\end{figure*}

We propose the steady state probability by using markov probability distribution as Prediction probability of steady state in fig . 7 is estimated as:

\begin{equation}
p(\pi_t)=\sum_{\pi_{t-1=1}}^Np(\pi_t|\pi_{t-1})p(\pi_{t-1})
\end{equation}

$P(\pi t)$ is probability of being at location $ \pi $ at time ``t``. $p(\pi_t|\pi_{t-1})$ is the probability of being at location $ \pi $ at time t given previous location $\pi$ at time $t-1$. This process search space to most likely region based on object motion.

Correction step\\
We modeled the observations corresponding to different pairs that received inside this cell as
\begin{equation}
p(\pi_t|V)= p(V|\pi_t)p(\pi_t)N
\end{equation}
$P(\pi_t|v)$ is the probability of being at location $ \pi$ at time t, given the RSSI value $ V $, received at time $ t $. $P(V|\pi_t)$ is the probability of having RSSI values $ V $ and $ p(\pi_t)$ is probability of being at that location (from prediction step). N is normalized factor.
In this technique, incresing the observation sequence lengths add more information and increase accuracy, however increase latency. HMM is more accurate location technique among all of them discussed in this section.
\subsection{GPS Based Location Tracking}
Global Positioning system (GPS) is considered most well known location tracking technique for outdoor environment. However, GPS is not available in many cell phones, consumes a lot of energy because direct line of sight to satellite is required. Extra chips for location tracking are required, thus increases expenses.
In [4], Network consists of mobile and static sensor nodes. Mobile nodes are equipped with GPS enabled devices. Mobile nodes moves in 3D space and periodically broadcast their position information through beacon messages. Static nodes receive beacon messages when they are in communication range of mobile nodes.
Static nodes calculate their position using equation of Sphere. In analytic geometry, a sphere with center (x0, y0, z0) and radius r is focus of all points (x, y, z). Beacon Overhead is defined as ``the average number of beacon messages transmitted during the total localization time'' and Beacon overheads is computed in [4] as:
\begin{equation}
\small
Beacon overhead =\frac{total Beacon Messages}{ total Mobile Sensor Nodes}
\end{equation}
In this scheme, although all mobile nodes continuously broadcast their location, first and last beacon messages received from the mobile node is recorded and other received beacon messages are discarded. Therefore,  beacon overheads is efficiently minimized.

\section{Conclusion}
In this paper, we compared several localization schemes for indoor and outdoor environments in WBASNs. Our results shows that particle filtering performed well in nonlinear systems for indoor environments. We proposed mathematical equations to determine location accuracy in terms of euclidean distance for deterministic finger printing location tracking scheme. We suggest mathematical equations for steady state probability of HMM location tracking scheme, we estimate that HMM outperforms all in terms of location accuracy and computational overheads.

\end{document}